\documentclass[3p,twocolumn]{elsarticle}

\usepackage{graphicx}
\usepackage{amssymb}
\usepackage{amstext}
\usepackage{rotating}
\usepackage{array}
\usepackage{nicefrac}

\usepackage{tabularx}
\newcolumntype{C}[1]{>{\centering\arraybackslash}p{#1}} % zentrierte Spalten mit Breitenangabe 
\newcolumntype{R}[1]{>{\raggedleft\arraybackslash}p{#1}} % rechtsbündig mit Breitenangabe 

\biboptions{square,numbers,sort&compress}

\journal{Journal of Crystal Growth}

\newcommand\degC{\ensuremath^{\,\circ}\text{C}}
\newcommand\pOO{\ensuremath p_{\text{O}_2}}

\begin{document}

\begin{frontmatter}



\title{On the effect of oxygen partial pressure on the chromium distribution coefficient in melt-grown ruby crystals}

\author{Steffen Ganschow\fnref{sg}}
\author{Detlef Klimm}
\author{Rainer Bertram}

\fntext[sg]{Corresponding author, e-mail address: steffen.ganschow@ikz-berlin.de}

\address{Leibniz-Institut f\"ur Kristallz\"uchtung, Max-Born-Stra\ss e 2, D-12489 Berlin, Germany}

\begin{abstract}
Small ruby crystals were grown by the Czochralski technique in different atmospheres and their actual chromium content was analysed by a wet chemical method. The chromium distribution coefficient $k$ was found to be strongly dependent on oxygen partial pressure $\pOO$. It ranges from $k=0.3$ in a reducing atmosphere to $k=1.2$ in a slightly oxidizing atmosphere and to a good approximation $k$ depends linearly on $\log\pOO$. The experimental data are discussed on the basis of thermodynamic equilibrium calculations.
\end{abstract}


\end{frontmatter}

% \linenumbers

%-------------------------------------------------
\section{Introduction}

Various techniques have been used to grow ruby single crystals including the Czochralski~\cite{OHara1968,Kvapil1974a,Kvapil1978,Perner1981}, Verneuil~\cite{Kvapil1974a,Adamski1968,Grabmaier1969,Seifert1972} and other melt growth techniques~\cite{Eickhoff1969,Saito1986,Song2005,Maier2007}, the hydrothermal method~\cite{Nguyen1972,Weirauch1973}, high-temperature solution (flux) growth~\cite{Nelson1964,Linares1965,Adams1966,Leonyuk2005}, and growth from the gas phase~\cite{OConnor1966}. However, reports on chromium segregation are only rarely found in the literature and the values of the distribution coefficient given by different groups vary in a wide range~\cite{Kvapil1974a,Song2005,Dils1962,Maier2007}. For the Czochralski and Verneuil techniques it was found~\cite{Kvapil1974a} that the chromium distribution coefficient drastically changes with growth atmosphere, from less than 1 in a virtually oxygen-free atmosphere to greater than 1 in an oxygen containing atmosphere. It is suggested that this effect is caused by changes in the chromium valence. 

This study aims to fill -- at least partially -- the gap in the quantitative description of the dependence of the chromium distribution coefficient on oxygen partial pressure for the technologically most important case of melt growth of ruby.

%-------------------------------------------------
\section{Experimental}

Small single crystals were grown by the Czochralski technique. 120~g of pre-melted Al$_{2}$O$_{3}$ granules (\emph{Spolchemie}, purity better 4N) with small amounts of Cr$_{2}$O$_{3}$ (\emph{Alfa Aesar}, 4N7) were melted in an inductively heated iridium crucible of 38~mm inner diameter. The total chromium content of the initial melt was $0.54\pm 0.01$~wt\%. The crystals were grown on $\vec{a}$-axis oriented Al$_{2}$O$_{3}$ seeds with a constant growth rate of 2.0~mm/h. After the growth was completed the crystals were extracted from the melt and cooled to room temperature in 9~hours. The experiments were carried out in different atmospheres yielding different oxygen partial pressures at the working temperature (i.e. melting temperature of Al$_{2}$O$_{3}$, cf. Table~\ref{tab:data}). Small crystals with maximum diameter of 8~mm and masses not exceeding 2.5~g each, of circular to elliptical cross section and dark wine-red to pink color were obtained. 

Chemical element analysis was carried out using an inductively coupled plasma optical emission spectrometer (ICP-OES) IRIS Intrepid HR Duo (\emph{Thermo Electron}). The spectrometer was calibrated with simple synthetic solution standards. Samples were ball-milled and analysed after microwave digestion in a mixture of phosphoric and sulfuric acid at 250$\degC$. The relative standard deviations (RSD) for the analysis were of the order of 2\%.  

Polarized transmission spectra were recorded using a \emph{Perkin Elmer} Lambda 19 spectrophotometer. The measured samples were ca.~0.4~mm thick slices cut perpendicular to the growth direction with both sides polished. The beam was polarized by a Glan-Taylor prism, and the transmitted light was detected by a photomultiplier inside an integrating sphere. The transmission spectra were corrected for reflection losses at the sample front and back surfaces. The wavelength-dependent reflection factor was calculated from the Fresnel equations for the case of normal incidence using tabulated data for the refractive index of corundum~\cite{Tropf1998}. From the obtained absorption coefficients at the maximum at 400~nm and around 550~nm for both polarisation states the chromium concentration was calculated using the reported absorption cross-sections~\cite{Cronemeyer1966}.

%-------------------------------------------------
\section{Results \& discussion}

Upon crystallization from a finite nutrient, segregation causes a continuous shift of melt composition and therewith a compositional gradient in the growing crystal.  If the crystallized fraction $g$ of the melt is small then the concentration of solute in the melt does not deviate substantially from its initial value and the distribution coefficient $k$ of the solute -- in this case chromium -- can be simply calculated as 
\begin{equation}
k=\frac{[\text{Cr}]_{\text{S}}}{[\text{Cr}]_{\text{initial melt}}}.
\label{eq:definition_k}
\end{equation}
where $[\text{Cr}]_{\text{S}}$ means the concentration (weight fraction) of chromium in the crystal. In Table~\ref{tab:data} the measured chromium concentrations are listed along with the calculated chromium distribution coefficient. In a slightly oxidizing atmosphere of N$_{2}$ with 10~vol\% CO$_{2}$ chromium is preferentially incorporated into the growing crystal ($k>1$). This observation is consistent with the Al$_{2}$O$_{3}$--Cr$_{2}$O$_{3}$ phase diagram~\cite[Fig. 00309]{Levin2004} comprising a significantly higher melting temperature of Cr$_{2}$O$_{3}$, and with other experimental observations~\cite{Grabmaier1969,Dils1962}. In a moderately reducing atmosphere, like e.g. a diluted mixture of CO and CO$_{2}$ in a volume ratio 4:1, segregation is reversed. Less chromium is incorporated than contained in the melt ($k<1$). This behavior conforms with the predicted reduction of the melting temperature of chromium oxide containing considerable amounts of suboxides~\cite[Fig. 00006]{Levin2004} and with crystal growth experiments conducted in reducing atmosphere~\cite{Song2005}.

Notably, the chromium concentration determined by optical transmission spectroscopy is always lower than that determined by wet chemical analysis and the relative deviation is larger for low oxygen partial pressure. This result seems to be reasonable since at low oxygen partial pressure a larger fraction of chromium ions occurs in the +2 state that does not contribute to the two absorption bands used for the concentration measurement. ICP measurement is insensitive to the oxidation state and results in a more reliable measure for the total chromium concentration and was therefore used for calculation of the distribution coefficient. 

\begin{table}
\begin{centering}
\begin{sideways}
\begin{tabular}{p{7cm}p{2.5cm}p{1.5cm}p{1.5cm}p{1.5cm}}
\hline 
composition (volume fraction) & $\pOO$ / bar & \multicolumn{2}{l}{$[\text{Cr}]_{\text{S}}$ in wt\%} & $k$\tabularnewline
 &  & ICP & VIS & \tabularnewline
\hline
90\% N$_{2}$ + 10\% CO$_{2}$ & $5.7\times10^{-2}$ & 0.84 & 0.66 & 1.55\tabularnewline
90\% Ar + 8\% CO + 2\% CO$_{2}$ & $1.1\times10^{-5}$ & 0.45 & 0.35 & 0.83\tabularnewline
N$_{2}$ & $\approx2\times10^{-6}$ & 0.42 & 0.29 & 0.77\tabularnewline
94.4\% Ar + 5\% H$_{2}$ + 0.6\% H$_{2}$O & $8.3\times10^{-8}$ & 0.17 & 0.096 & 0.32\tabularnewline
\hline
\end{tabular}
\end{sideways}
\par\end{centering}
\caption{Atmospheres used in the growth experiments and resulting oxygen partial pressure $\pOO$ at 2053$\degC$. Chromium concentration in the grown crystals determined by wet chemical analysis (ICP) and transmission spectroscopy (VIS). For the calculation of the distribution coefficient $k$ the ICP results were used.}
\label{tab:data}
\end{table}

\begin{figure}
\begin{centering}
\includegraphics[width=\columnwidth]{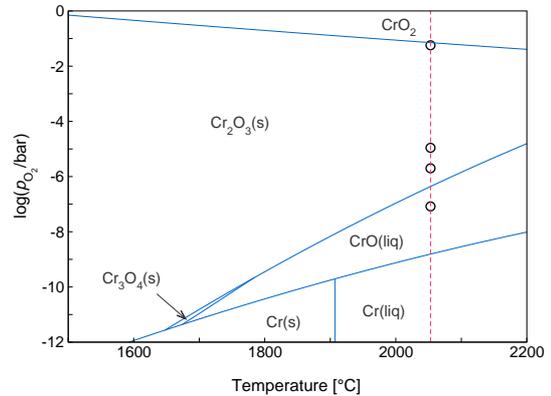}
\par\end{centering}
\caption{Calculated predominance diagram of the Cr--O system. The red dashed line stands for the melting temperature of Al$_{2}$O$_{3}$; the circles mark oxygen partial pressures applied in the experiments.
\label{fig:predom}}
\end{figure}

Like other transition metals, chromium can take on various oxidation states depending on the ambient oxygen partial pressure $\pOO$. The most common oxide is chromium(III) oxide, Cr$_{2}$O$_{3}$. Its stability field ranges over several orders of magnitude oxygen partial pressure from room temperature up to 2000$\degC$. At temperatures above 2000$\degC$ other chromium oxides may emerge -- chromium(IV) oxide CrO$_{2}$ at high oxygen partial pressure and chromium(II) oxide CrO at low oxygen partial pressure. Chemical equilibrium between different oxides can be described by the reaction
\begin{equation}
2\,\text{CrO}_{m/2}\,+\,\nicefrac{1}{2}\,\text{O}_{2}\,\rightleftharpoons\,2\,\text{CrO}_{(m+1)/2}\label{eq:reaction}
\end{equation}
with $m$ being the chromium valence. As long as oxygen is the only gaseous phase involved, i.e.\ as long as the vapour pressure of the condensed phases is low compared to the oxygen pressure, the predominance (Ellingham) diagram of the system (Fig.~\ref{fig:predom}), consisting of plots of the equilibrium oxygen fugacity versus temperature for reactions (\ref{eq:reaction}), can be easily computed from tabulated thermodynamic data~\cite{Pelton1991,Klimm2009}. The diagram area is divided into fields of predominance of a certain oxide, which is the oxide that will be formed because its formation is associated with the  reduction of Gibbs energy of the system. 

In the experimental $T$--$\pOO$ domain of the Al--Cr--O system the only stable aluminium oxide is Al$_{2}$O$_{3}$ whereas four different chromium oxides, namely CrO$_{2}$, Cr$_{2}$O$_{3}$, CrO, and Cr$_{3}$O$_{4}$, and chromium metal Cr may occur. At the  melting temperature of Al$_{2}$O$_{3}$ ($T^{f}=2053\degC$, vertical line in Fig.~\ref{fig:predom}) the range of oxygen partial pressure that stabilizes the trivalent chromium ion Cr$^{3+}$ is fairly wide and can be easily met in practical experiments. This fact explains why ruby can be grown in reasonable quality without great care for the growth atmosphere. 

If, however, the oxygen partial pressure is low, as it is the case for most ``inert'' atmospheres used to prevent the metallic crucible from oxidation, Cr$^{2+}$ may be formed. In the case that the amount of Cr$^{2+}$ ions becomes significant, crystallization cannot be explained by the Al$_{2}$O$_{3}$--Cr$_{2}$O$_{3}$ phase diagram that assumes only trivalent chromium. Instead, discussion must be extended to a system allowing also oxygen pressure to be variable. For the purpose of this study, the system Al$_{2}$O$_{3}$--Cr$_{2}$O$_{3}$--CrO seems to be appropriate unless $\pOO$ is so low that Cr metal may be formed.

\begin{figure}
\begin{centering}
\includegraphics[width=\columnwidth]{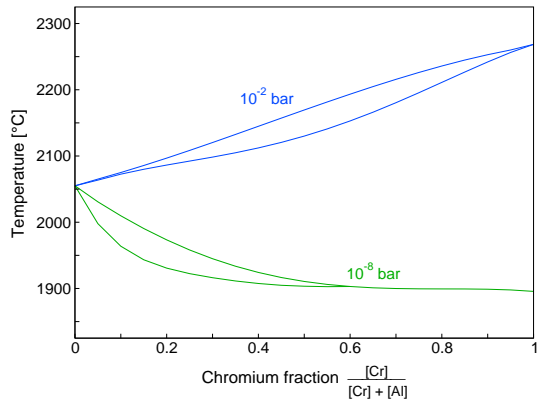}
\par\end{centering}
\caption{Calculated sections through the CrO--Cr$_{2}$O$_{3}$--Al$_{2}$O$_{3}$ system for different oxygen partial pressures $\pOO$.  At low $\pOO$ addition of chromium oxide to the Al$_{2}$O$_{3}$ melt lowers melting temperature and consequently $k<1$ (red dashed curve). In contrast, at high $\pOO$ (blue solid curve) melting temperature is raised and $k>1$. 
\label{fig:section}}
\end{figure}

Available thermodynamic data in the CrO--Cr$_{2}$O$_{3}$--Al$_{2}$O$_{3}$ system including those of liquid and solid solutions permit calculation of phase equilibria for arbitrary $(T,\pOO)$ conditions using numerical algorithms for Gibbs energy minimization~\cite{Pelton1991,FactSage,Degterov1996}. In particular, sections of constant oxygen fugacity may be computed~(Fig.~\ref{fig:section}) that allow one to estimate the chromium equilibrium distribution coefficient $k_{0}$ as the ratio of the solid and liquid Cr$_2$O$_3$ mole fraction, $k_{0}=x_\text{S}/x_\text{L}$, at a temperature corresponding to $x_\text{L}\approx 0.05$.
It should be noted that these sections are not necessarily binary ones. In general the solid in equilibrium with a liquid of given composition will have a different oxygen fugacity and will therefore lie off the section. If these deviations are small this approximation will still be reasonable. 

\begin{figure}
\begin{centering}
\includegraphics[width=\columnwidth]{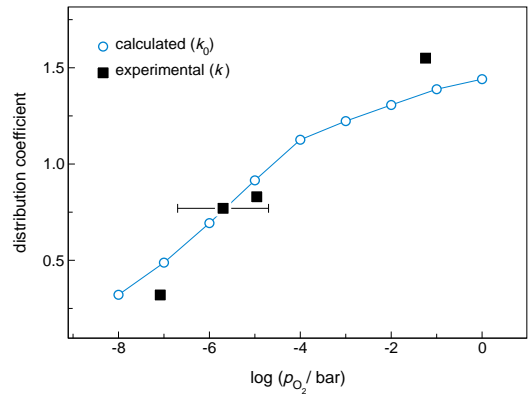}
\par\end{centering}
\caption{Chromium distribution coefficient versus oxygen partial pressure. The error bars indicate the estimated inaccuracy in $\log p_{\text{O}_{2}}$ for gas mixtures not containing any oxygen-bearing constituent for which oxygen partial pressure arises from residual impurities of the gases and/or moisture adsorbed e.g. at the chamber walls and insulation.
\label{fig:segcoeff}}
\end{figure}

The calculated dependence of $k_{0}$ on $p_{\text{O}_{2}}$ is shown in Fig.~\ref{fig:segcoeff}. In good quantitative agreement with the experimental results, $k_{0}>1$ for moderate oxygen pressure, and $k_{0}<1$ for very low oxygen pressure. Similar observation was made by Kvapil et al.~\cite{Kvapil1974a}. In their Czochralski grown crystals they found a distribution coefficient ranging between 0.2 in a reducing atmosphere (Ar with 2\% H$_{2}$), and a maximum of 1.2 for a slightly oxidizing atmosphere (N$_{2}$ with 1...2\% O$_{2}$% 
\footnote{In the original work a value of 0.01...0.02\% is given. However, an admixture of a very few per cent oxygen to an inert gas is commonly used to achieve a slightly oxidizing atmosphere that prevents selected oxides, e.g. Ga$_{2}$O$_{3}$, from being reduced. Therefore we suppose that the correct value is 1...2\%.}). 
Quantitative comparison of the {\em effective} distribution coefficient obtained experimentally with the {\em equilibrium} distribution coefficient derived from thermodynamics is possible because according to the theory of Burton, Prim, and Slichter~\cite{Burton1953} for the given growth conditions, i.e.\ low growth rate and assumed complete mixing in the liquid, the difference between both is negligible.

Reversal of the segregation behavior occurs at $\pOO\approx10^{-4}$~bar. The point [$T=2050\degC;\;\pOO=10^{-4}\:\text{bar}$] lies well inside the predominance region of Cr$_{2}$O$_{3}$~(Fig.~\ref{fig:predom}). Therefore it can be assumed that under these $(T,\pOO)$ conditions chromium occurs almost exclusively as Cr$^{3+}$ that stabilizes the corundum structure (cf. the higher melting temperatures of Al$_{2}$O$_{3}$ and Cr$_{2}$O$_{3}$) and is therefore preferentially incorporated into the growing crystal. If oxygen pressure is reduced below 10$^{-4}$~bar, an increasing part of the chromium is reduced to Cr$^{2+}$. The divalent ion is significantly larger than the Al$^{3+}$ host ion (for sixfold coordination $r_{\text{Al}^{3+}}=67.5\:\text{pm},$ $r_{\text{Cr}^{2+}}=87\: \text{pm}$~\cite{Shannon1976}) and, when incorporated, would largely destabilize the corundum structure. It is therefore rejected from the interface and lowers the measured chromium distribution coefficient to a degree increasing with its concentration in the melt. The gradual concentration-dependent effect is also reflected by the monotonic shape of the $k_{0}$~vs.~$\log p_{\text{O}_{2}}$ curve obtained in thermodynamic calculations. In this model the slope of the curve changes around $\log\pOO=-4$, i.e. at the oxygen partial pressure for which segregation reverses.

%-------------------------------------------------
\section{Conclusion}

The experimentally observed dependence of the chromium distribution coefficient on oxygen partial pressure can be reproduced by thermodynamic equilibrium calculations within the CrO--Cr$_{2}$O$_{3}$--Al$_{2}$O$_{3}$ system. A strongly reducing atmosphere increases the amount of Cr$^{2+}$ ions in the melt that, in turn, lower the melting temperature of solid solutions and the observed chromium distribution coefficient therein. Quantitative knowledge of this dependence allows identification of an oxygen partial pressure for which segregation will be practically eliminated. For ruby, this can be expected to occur for $\pOO\approx10^{-4}\:\text{bar}$. Melt growth in this pressure range promises to yield crystals of homogeneous chromium distribution. If necessary, subsequent annealing in an appropriate atmosphere may be used to re-charge Cr$^{2+}$ ions. This procedure has the potential to be applied to other systems containing multivalent ions, e.g. transition metals but also some of the rare earths. 

The concept of the effective distribution coefficient introduced by Burton, Prim, and Slichter~\cite{Burton1953} shows a way to affect solute distribution by adjusting growth conditions, but there is an intrinsic limitation. The value of the effective distribution coefficient $k$ always lies between that of the equilibrium distribution coefficient $k_0$ and unity, $0 \leq |k -1| \leq |k_0 - 1|$. In contrast, the current approach manipulates the type of species that are available in front of the growth interface. Upon incorporation, the diverse species may behave differently. In the case of ruby considered here, one species may be preferably incorporated ($k_0>1$) while another may be rejected by the preceding interface ($k_0<1$). This fact allows design of  growth processes that overcome the limitations imposed on the effective distribution coefficient. The approach acts on the level of liquid--solid phase diagrams and is therefore relevant to all techniques using this phase transition, i.e.\ to all melt growth techniques.

%-------------------------------------------------
\section*{Acknowledgements}

The authors gratefully acknowledge the technical assistance of Mario Br\"utzam and Mike Pietsch.



%% References
%%   \cite{key}         ==>>  Author [#]


\section*{References}

\bibliographystyle{elsarticle-num}
\bibliography{literature}






% \end{thebibliography}

\end{document}